\documentclass{mn2e}

\renewcommand\[{\begin{equation}}
\renewcommand\]{\end{equation}}

\def\d{{\rm d}}
\def\p{\partial}
\def\i{\relax\ifmmode{\rm i}\else\char16\fi}

\def\lesssim{{_ <\atop{^\sim}}}

\def\b#1{{\bf{#1}}}

\def\lesssim{\mathrel{\hbox{\rlap{\hbox{\lower4pt\hbox{$\sim$}}}\hbox{$<$}}}}
\def\gtrsim{\mathrel{\hbox{\rlap{\hbox{\lower4pt\hbox{$\sim$}}}\hbox{$>$}}}}

\def\apj#1 #2{ApJ, #1, #2}
\def\aj#1 #2{AJ, #1, #2}
\def\mn#1 #2{MNRAS, #1, #2}
\def\aa#1 #2{A\&A, #1, #2}

\begin{document}

   \title[advection dominated accretion]
   {On the impossibility of advection dominated accretion}

   \author[J. Binney]
          {James Binney
           \\
           Theoretical Physics, Department of Physics, 1 Keble Road, Oxford OX1 3NP\\
          }

   \date{}

   \maketitle

\begin{abstract}
Using only the assumption that all interactions between  particles
in an accretion flow are electromagnetically mediated, it is shown that the
time to establish equipartition between ions and electrons is shorter than
the characteristic accretion time. Consequently, two-temperature fits to the
spectra of accreting objects are unphysical, and models in which significant
thermal energy is carried across the event horizon are effectively ruled out.
\end{abstract}


\section{Introduction}

The supermassive black holes at the centres of galaxies are usually much
less luminous than simple accretion theory would predict: the black hole is
surrounded by thermal plasma whose properties may be known from X-ray
observations, and the black hole should accrete this plasma at roughly the
Bondi-Hoyle rate. When this accretion rate is combined with the radiative
efficiency $\sim0.1$ that follows from comparison of the integrated luminosity of
quasars with the observed space density of supermassive black holes (Yu \&
Tremaine, 2002), a luminosity is derived that generally exceeds that
observed by 3--5 orders of magnitude (Loewenstein et al.~2001; Di Matteo et
al 2001, 2003; Pellegrini et al 2003).

The low measured luminosities of nuclear black holes led to the development
of the advection-dominated accretion flow -- hereafter ADAF model -- in
which the energy released as plasma falls towards the black hole is retained
as internal energy within the plasma and ultimately carried over the black
hole's event horizon rather than escaping to infinity (Ichimaru 1977; Rees
et al.~1982; Narayan \& Yi, 1994, 1995; Igumenshchev, Chen \& Abramowicz,
1996; see also the lucid review in Chapter 11 of Frank, King \& Raine,
2002). These ideas were subsequently applied to accreting stellar-mass black
holes and neutron stars. The flows around these objects would be similar
down to radii comparable to the surface of the neutron star. At this radius
the advected energy would be radiated in the neutron star case, while in the
case of a black hole the energy would be carried on through the event
horizon. Hence the existence of an event horizon around a black hole would
cause accreting stellar-mass black holes to be much less luminous than
comparable neutron stars.  Garcia et al.~(2001) claim to observe this effect.

Since material reaching the event horizon has lost gravitational potential
energy that is a substantial fraction of $c^2$ per unit mass, the ADAF model
requires that the innermost plasma achieves temperatures of order
$m_pc^2\sim1\,$GeV. At such temperatures the electrons are highly
relativistic ($\gamma\sim1000$), the radiative efficiency of the plasma
would be expected to be high because the electrons' radiative losses
increase with temperature faster than $T_{\rm e}^7$ (Rees et al.~1982). In order to limit
the radiative efficiency of the plasma, the ADAF model conjectures that the
electrons decouple thermodynamically from the ions, which are the component
of the plasma that receives the lion's share of the released gravitational
energy.  Hence the thermal motions of the ions become mildly relativistic as
the event horizon is approached, while the electrons, which dominate
radiative processes, remain at temperatures that are orders of magnitude
lower.

Proponents of the ADAF model motivate the decoupling of the electrons from
the ions by pointing out that in a highly viscous low, the residence time of
any given electron or ion in the flow is small, so the plasma density
required to achieve a given accretion rate is low. At low densities and high
temperatures, the time required for coulomb scattering to establish
equipartition of energy between ions and electrons becomes long, and can
easily exceed the residence time (Rees et al.~1982).  In this paper I give a
simple argument of considerable generality, which implies that the time
$t_{\rm equi}$ required to establish equipartition between ions and
electrons is always shorter than the residence time $t_{\rm res}$.
Consequently, one may safely assume that the electrons are at the same
temperature as the ions.  This result invalidates most applications of the
ADAF model, and effectively rules out all models in which significant thermal
energy is carried across the event horizon.

\section{The equipartition time}

The plasma comprises particles that interact electro-magnetically with
each other and with a given gravitational field. I shall assume that the
dynamics can be treated in a non-relativistic approximation -- the
generalization of the analysis to the relativistic case is not hard, but
there is a loss of clarity and simplicity.

The Hamiltonian for motion of a particle of mass $m$ and charge $q$ in an
electromagnetic field that is characterized by the potentials $\psi$ and $\b
A$ is
 \[\label{defsH}
H={(\b p-q\b A)^2\over2m}+q\psi+m\Phi.
\]
 where $\Phi$ is the gravitational potential. From standard Hamiltonian
theory we know that
 \begin{eqnarray}\label{Hdot}
{\d H\over\d t}&=&{\p H\over\p t}
=-q{\b p-q\b A\over m}\cdot{\p\b A\over\p t}+q{\p\psi\over\p t}\nonumber\\
&=&-q\b v\cdot{\p\b A\over\p t}+q{\p\psi\over\p t},
\end{eqnarray}
 where we have assumed that $\Phi$ is time-independent and have identified
the particle velocity $\b v=(\b p-q\b A)/m$. 
Since the right side of this equation is proportional to the charge
$q$, if at some location energy is lost by one species, it is gained by the
oppositely charged species. Thus this equation describes the mechanism by
which equipartition is established between ions and electrons; the net
direction of the energy flow is mandated by the general principles of
statistical physics, and the rate of flow may be estimated from
equation  (\ref{Hdot}).

The two terms on the right of equation (\ref{Hdot}) have different physical
origins: the first describes work done by the inductive e.m.f.s associated
with time-dependent magnetic fields, while the second describes work done by
the electrostatic potential, during two-body scattering and in anchoring the
fast electrons to the gravitationally confined protons. In view of this
distinction, the two terms will not cancel with any exactitude, and we may
obtain a lower limit on the rate of energy transfer between species by
considering only one term.  For reasons that should become clear later, we
focus on the inductive term and obtain the following upper limit on the
equipartition time
 \[\label{tequi1}
t_{\rm equi}\sim{H\over|\d H/\d t|}<{H\over|q\b v\cdot\p\b A/\p t|}.
\]

Inward drift through
an accretion disk is controlled by loss of angular momentum $L_z$ . Again using
standard Hamiltonian theory together with the result that $L_z=p_\phi$ is
the momentum conjugate to the azimuthal angular coordinate $\phi$, we have
 \begin{eqnarray}\label{Ldot}
{\d L_z\over\d t}&=&[p_\phi,H]=-{\p H\over\p\phi}\nonumber\\
&=&q\b v\cdot{\p\b A\over\p\phi}-q{\p\psi\over\p\phi},
 \end{eqnarray}
 where the square bracket is a Poisson bracket.

The expressions above give the rates of energy and angular momentum change
for a single particle. To obtain the corresponding rates for the plasma as a
whole we have to sum over the particles in some volume. 
\begin{eqnarray}\label{Ltotdot}
{\d L_z^{\rm tot}\over\d t}
&=&{\d L_z^+\over\d t}+{\d L_z^-\over\d t}\nonumber\\
&=&\sum|q|\left((\b v^+-\b v^-)\cdot{\p\b A\over\p\phi}-{\p\psi\over\p t}\right)\\
&=&\int\d^3\b x\,\left(\b j\cdot{\p\b A\over\p\phi}-\rho{\p\psi\over\p\phi}\right),\nonumber
\end{eqnarray}
 where $\b j$ is the current density and $\rho$ is the charge density. 

If we again neglect the term containing $\psi$, we find that the residence
time within the flow is
 \[\label{tres}
t_{\rm res}\sim{L_z^{\rm tot}\over|\d L_z^{\rm tot}/\d t|}\sim
{L_z^{\rm tot}\over|\int\d^3\b x\,\b j\cdot\p\b A/\p\phi|}.
\]

We now return to our upper limit (\ref{tequi1}) on the equipartition time.
We average top and bottom over all particles in a given small region. This
operation is bound to increase the ratio significantly since the
Hamiltonian values on the top are all positive, while those on the bottom,
being proportional to $q$,
have a tendency to cancel between species. Consequently, our upper limit
remains an upper limit, and we may write
 \[\label{tequi2}
t_{\rm equi}<{H^{\rm tot}\over|\int\d^3\b x\,\b j\cdot\p\b A/\p t|}.
\]
 Dividing by equation (\ref{tres}) we obtain
 \[\label{tequitres}
{t_{\rm equi}\over t_{\rm res}}<{H^{\rm tot}\over L_z^{\rm tot}}
{|\int\d^3\b x\,\b j\cdot\p\b A/\p\phi|\over
|\int\d^3\b x\,\b j\cdot\p\b A/\p t|}.
\]
 Standard Hamiltonian theory tells us that for any particle the azimuthal
frequency $\Omega_\phi$ is given by $\Omega_\phi=\p H/\p L_z\sim H/L_z$, so
the first ratio on the right side of (\ref{tequitres}) can be approximated
by the local circular frequency. The ratio of integrals we approximate by
the ratio of the corresponding derivatives of $\b A$, so
 \[\label{tequitres2}
{t_{\rm equi}\over t_{\rm res}}<\Omega_\phi
{|\p\b A/\p\phi|\over|\p\b A/\p t|}.
\]
 To get an idea of the relative sizes of the derivatives of $\b A$, consider
the case in which $\b A(t,\phi)$ can be approximated by a pattern that
propagates at some phase frequency $\omega$. Then $A$ is a function of the
single variable $\omega t-\phi$ and the ratio of derivatives becomes
$1/\omega$.  The important case is that in which the pattern propagates with
some particles (the particles that are Landau-damping it). In this case
$\omega$ is of order $\Omega_\phi$ and the right side of
(\ref{tequitres2})  evaluates to unity. In principle the ratio
$\Omega_\phi/\omega$ could be equal to some small integer because the
Landau-damping resonance could be associated with a harmonic of $\omega$.
However, in view of the large amount by which the right side of
(\ref{tequi2}) exceeds that of (\ref{tequi1}) we may state with confidence
that 
 \[\label{atlast}
{t_{\rm equi}\over t_{\rm res}}<1.
\]

\section{Discussion}

In view of this simple and very general demonstration that the time to
establish equipartition between electrons and ions is not greater (and
realistically much smaller) than the time it takes plasma to move through
the accretion disk, why has so much attention has been paid to the proposal
that the electron and ion temperatures differ significantly? The answer lies
in uncertainty as to what mechanism actually drives accretion through a
disk. It was very early on realised that the viscosity provided by standard
kinetic theory falls short of that required by many orders of magnitude
(e.g., Frank et al.~2002). In
default of a proper physical theory, the field progressed by using the
$\alpha$ parameter of Shakura \& Sunyaev (1973) to connect viscosity to the
pressure in a dimensional sense. The ADAF model assumed that a large value
of $\alpha$ was appropriate, and thus that the viscosity was very much
larger than kinetic theory would predict. Inconsistently, it assumed that
the equipartition time was still correctly estimated by kinetic theory. All
the derivation above does is to insist that whatever viscosity the disk
experiences is mediated by electromagnetic fields, and to show that these
fields inevitably establish equipartition on a timescale that is shorter
than the accretion timescale.

Since the seminal paper of Balbus \& Hawley (1991) it has become widely
agreed that the magnetic-rotational instability (MRI) is the detailed
mechanism that provides the high values of viscosity that observations
imply. In this picture the electromagnetic field in the disk is dominated by
time-varying magnetic fields with scale lengths comparable to the disk
thickness. The derivation above implicitly focuses on this case in as much
as it neglects the terms in (\ref{Hdot}) and (\ref{Ltotdot}) that contain
the electrostatic potential $\psi$. If one were concerned about
kinetic-theory viscosity, one would need to include these terms. It is easy
to see that in the limit in which these terms dominate the terms in $\b A$,
one would again arrive at our fundamental result (\ref{atlast}).

\section{summary}

A time-dependent electromagnetic field transfers energy between oppositely
charged particles very efficiently. Consequently, when one compares the rate
of such energy transfer with the rate at which the field mediates the
exchanges of angular momentum that drive accretion, one concludes that
equipartition of energy between electrons and ions can be achieved well
within the time it takes plasma to drift inwards. Hence we may state with
confidence that $T_{\rm ion}=T_{\rm electron}$.
Fits of the observed spectra of sources such as Sgr A$^*$ to the ADAF model
(Narayan \& Yi, 1995) assume that $T_{\rm ion}\gg T_{\rm electron}$. Such
fits are unphysical.

In an $\alpha$ disk the density scales with the mass-flow rate as $\dot
m^{11/20}$, while the residence time $t_{\rm res}$ scales as $\dot
m^{-3/10}$ (Frank et al.~2002). Since the time $t_{\rm rad}$ required to radiate a
given energy per particle scales inversely with density, the ratio $t_{\rm
rad}/t_{\rm res}$ scales as $\dot m^{-1/4}$, and we conclude that at
sufficiently small values of $\dot m$, even thermalized electrons will be
unable to radiate the internal energy of a plasma that is at the virial
temperature before the event horizon is reached. Hence the ADAF model is not
completely excluded.  However, for an ADAF to be possible, the ions must
become relativistic, and, in view of the efficiency of equipartition, the
electrons must become ultrarelativistic.  Such electrons radiate extremely
efficiently through a combination of bremsstrahlung, synchrotron radiation
and the Compton scattering of photons.  Consequently, the values of $\dot m$
at which radiative cooling of the electrons is unimportant are too low to be
of astrophysical interest (Rees et al.~1982).

The nature of accretion onto black holes remains a puzzle. There is
considerable observational evidence that energy released by accretion onto
a black hole can be channelled into a bipolar outflow with remarkably little
waste heat being radiated (e.g., Churazov et al.~2003). The ADAF problem was
developed to explain this fact, and if the ADAF model must be rejected,
another explanation must be sought.

 The evidence for highly efficient generation of outflows is clearest when
jets with bulk Lorentz factors of a few are observed. It seems likely that
these jets emerge from the stressed vacuum around the black hole (Blandford
\& Znajek 1977) and do not address the issue of how accretion flows work. It
is widely believed that outflows, presumably at the local Kepler speed, are
an integral part of accretion-flow dynamics as in the ADIOS model of
Blandford \& Begelman (1999). However, it remains very unclear how these
outflows avoid an unacceptable level of radiative dissipation from the
underlying accretion flow.

\section*{acknowledgements}
I am grateful to Andrew King for commenting on the first draft of this
paper.

\end{document}